# NeRP: Implicit Neural Representation Learning with Prior Embedding for Sparsely Sampled Image Reconstruction

Liyue Shen, John Pauly, Lei Xing

*Abstract*—**Image reconstruction is an inverse problem that solves for a computational image based on sampled sensor measurement. Sparsely sampled image reconstruction poses addition challenges due to limited measurements. In this work, we propose an implicit Neural Representation learning methodology with Prior embedding (NeRP) to reconstruct a computational image from sparsely sampled measurements. The method differs fundamentally from previous deep learning-based image reconstruction approaches in that NeRP exploits the internal information in an image prior, and the physics of the sparsely sampled measurements to produce a representation of the unknown subject. No large-scale data is required to train the NeRP except for a prior image and sparsely sampled measurements. In addition, we demonstrate that NeRP is a general methodology that generalizes to different imaging modalities such as CT and MRI. We also show that NeRP can robustly capture the subtle yet significant image changes required for assessing tumor progression. Code: *https://github.com/liyues/NeRP*.**

*Index Terms*—**implicit neural representation, prior embedding, sparsely sampled image reconstruction, inverse problem,**

## I. Introduction

IMAGE reconstruction is conventionally formulated as an inverse problem, with the goal of obtaining the computational image of an unknown subject from measured sensor data. For example, projection data are measured for computed tomography imaging (CT) while frequency domain (k-space) data are sampled for magnetic resonance imaging (MRI). To reconstruct artifact-free images, dense sampling in measurement space is required to satisfy the Shannon-Nyquist theorem. However, for many practical reasons, such as imaging dose reduction in CT imaging and acceleration of MRI, it is highly desirable to reconstruct images from sparsely sampled data. The ill-posed nature of the sparse sampling image reconstruction problem poses a major challenge for algorithm development. Many approaches have been studied to solve this problem. One widely used approach is to exploit prior knowledge of the sparsity of the image in a transform domain, such as in compressed sensing, where total-variation, low-rank, and dictionary learning have been applied [1]-[7].

Clinically, the patients often need to be scanned serially over time. For example, in cancer treatment, the patients are scanned before and after therapy or even during the therapy, to plan the treatment, guide the interventions, and monitor the therapeutic response. In radiation therapy, for example, the treatment often involves multiple fractions (sometimes, the number of fractions can be more than 30). In each fraction, the patient is scanned before or even during treatment for patient setup and tumor target localization. Sparse sampled image reconstruction is imperative to speed up the imaging process for efficient clinical workflow and for nearly real-time tumor target localization in image guided interventions. Additionally, sparse imaging also allows radiation dose reduction in CT imaging. These motivate us to search for an effective way to take advantage of previously scanned images.

Unprecedented advances in deep learning driven by learning from large-scale data have achieved impressive progress in many fields, including computational image reconstruction. Many researchers have introduced deep learning models for medical imaging modalities such as CT and MRI [8], [9]. The key to these deep learning approaches is training convolutional neural networks (CNNs) to learn the mapping from raw measurement data to the reconstructed image by exploiting the large-scale training data as shown in Fig.1(a). The network exploits the hidden transformation information embedded in the data through the data-driven training procedure. More advanced methods have improved the conventional deep learning reconstruction by leveraging prior knowledge from other aspects, such as generative adversarial model [10], [11] and geometry-integrated deep learning frameworks [12]-[15]. Previous works have demonstrated the effectiveness of explicitly incorporating physics and geometry priors of imaging system with deep learning [12], [15].

Although these works show the advantage of deep learning for medical image reconstruction, they have also exposed some limitations. For example, the acquisition of large-scale training datasets can be a bottleneck, the reconstructions may not be robust when deployed to unseen subjects, the reconstructions

This work was supported in part by Stanford Bio-X Graduate Student Fellowship and NIH/NCI 1R01CA227713 and 1R01CA256890.

Liyue Shen is with the Electrical Engineering Department, Stanford University, Stanford, CA 94305 USA (e-mail: liyues@stanford.edu).

John Pauly is with the Electrical Engineering Department, Stanford University, Stanford, CA 94305 USA (e-mail: pauly@stanford.edu).

Lei Xing is with the Radiation Oncology Department and Electrical Engineering Department, Stanford University, Stanford, CA 94305 USA (e-mail: lei@stanford.edu).

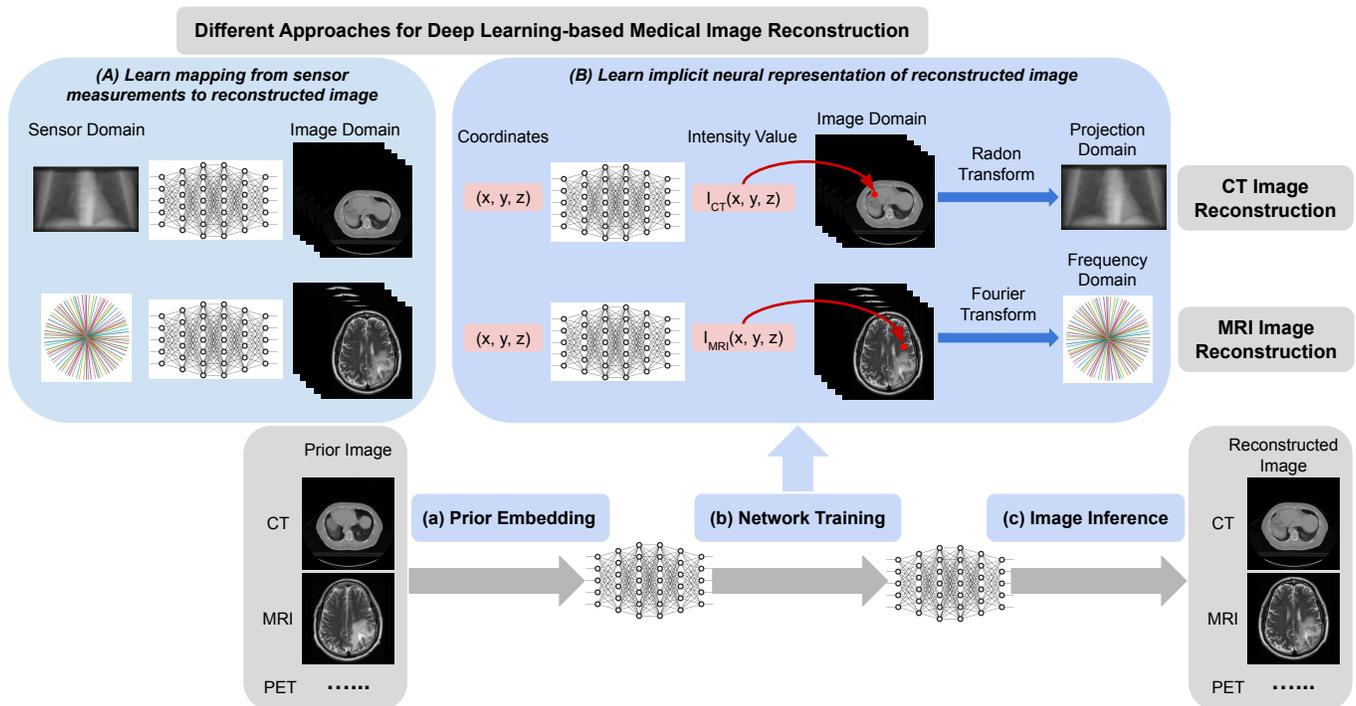

Fig. 1. Schematic illustration of different approaches for deep learning-based medical image reconstruction. (A) Deep neural network is developed to learn the mapping from the sensor (measurements) domain (e.g. projection space, frequency space) to image domain (e.g. CT, MRI). (B) Deep neural network is developed to learn the implicit neural representation of the reconstructed image. The input to the network is the spatial coordinates for any points within the image field, while the output is the corresponding intensity values. Any image (e.g. CT, MRI) can be implicitly represented by a continuous function which is encoded into the neural network's parameters. The proposed image reconstruction framework of implicit neural representation learning with prior embedding (NeRP) consists of three modules: (a) Prior Embedding: prior image is embedded into the neural network's parameters by training with coordinate-intensity pairs. (b) Network Training: using the prior-embedded network as initialization, we train the reconstruction network constrained by the sparse samples in sensor (measurement) domain (i.e. projection domain for CT imaging, frequency domain for MRI imaging), and learn the implicit neural representation for the reconstructed image. (c) Image Inference: the reconstructed image is obtained by inferring the trained network across all the spatial coordinates in the image grid.

can be unstable with subtle yet significant structural changes such as tumor growth, and there can also be difficulties generalizing to different image modalities or anatomical sites [16]. To address these limitations, we propose a new deep learning methodology for image reconstruction. We propose to learn the implicit Neural Representation of an image with Prior embedding (NeRP), instead of learning the reconstruction mapping itself. This is rather different from previous deep learning-based reconstruction methods. Leveraging the new insight of neural representation learning, we also present a succinct and effective way to embed the longitudinal prior image as the network parameters. This provides a systematic framework to incorporate the longitudinal prior knowledge into deep learning, which is useful to various medical imaging applications such as image guided interventions.

The NeRP idea is illustrated in Fig.1. Conventionally, a neural network is trained to learn the mapping from the sampled measurement data to reconstruct images based on a large-scale training database. The proposed NeRP model learns the network, i.e. multi-layer perceptron (MLP), to map the image spatial coordinates to the corresponding intensity values. The neural network learns the continuous implicit neural representation of the entire image by encoding the full image spatial field into the weights of MLP model [17], [18]. Image reconstruction is then reduced to simply querying the model embedded in the network. The image reconstruction problem is transformed into a network optimization problem. Instead of searching for the best match in the image space, we reconstruct the image by searching for it in the space of networks' weights.

The network is trained to match the subsampled measurements in raw data space (e.g. projection space sampling for CT or frequency space sampling for MRI), by integrating the forward model of the corresponding imaging system. For sparse sampling, the measurements may not provide sufficient information to precisely reconstruct images of the unknown subject due to the ill-posed nature of the inverse problem. The proposed NeRP framework exploits prior knowledge from a longitudinal previous image of the same subject. This is particularly applicable to clinical protocols, where patients are scanned multiple times during disease management for the purpose of guiding the interventional procedure or monitoring the therapeutic response. In this situation, the previously scanned image provides valuable information of the patient anatomy and can be utilized as the prior knowledge for subsequent imaging of the patient.

The implicit neural representation first embeds the internal information of the prior image into the weights of MLP. This serves as the initialization for the search for the representation of the target image. Starting from this prior-embedded initialization, the network can be optimized to reach an optimal point in the function space given only sparsely sampled measurements. Finally, the learned MLP can generate the image reconstruction by traversing all the spatial coordinates in the image space. Note that NeRP requires no training data form external subjects except for the sparsely sampled measurements and a prior image of the subject.

The main contributions of this work are:

1) We present a novel deep learning methodology for sparsely sampled medical image reconstruction by learning the implicit neural representation of image with prior embedding (NeRP). By reformulate the image reconstruction problem as an optimization of a continuous function, our method alleviates the need for a large amount of population-based training data from external subjects, and can be easily generalized across different imaging modalities and contrasts, and different anatomical sites.

2) Based on the neural representation learning, we propose a succinct and effective way to incorporate the longitudinal prior knowledge into the deep learning by embedding the longitudinal prior image as the network parameters, which makes it possible to incorporate the longitudinal information into deep learning models for various applications.

3) We present extensive experiments for both 2D and 3D image reconstruction with various imaging modalities, including CT and MRI, and demonstrate the effectiveness and generalizability of the proposed NeRP method. In particular, we show that our method is robust for capturing subtle yet significant structural changes such as those due to tumor progression. Overall, the salient features of the proposed framework lay a solid foundation for the potential clinical applications.

## II. RELATED WORK

### A. Deep Learning for Medical Image Reconstruction

Most previous work in deep learning-based medical image reconstruction trains the deep neural networks to learn the mapping from sampled measurements to reconstructed images [8]-[11]. Zhu *et al.* [8] introduced a CNN model for manifold learning and mapping between the sensor and image domain for accelerating the acquisition and reconstruction in MRI. Meanwhile, Shen *et al.* [9] proposed a deep network with a transformation module in high-dimensional feature space to learn the mapping from 2D projection radiographs to 3D anatomy for reconstructing volumetric CT images from sparse views. Later, adversarial learning was introduced to train the CNN model for obtaining better reconstruction image quality [10], [11]. Mardani *et al.* [10] used generative adversarial networks (GAN) to model the low-dimensional manifold of MR images. Ying *et al.* [11] adopted the adversarial loss in the network training objective for 2D-3D CT reconstruction. Although GAN-based methods can reconstruct more realistic images with higher image quality, the generated fine details and structures may not be reliable due to the synthetic nature of GAN models [11].

Recently, human prior knowledge has been considered to guide the construction of deep learning models [12]-[15]. For example, geometry-integrated deep learning methods were proposed for CT image reconstruction [12] and X-ray projection synthesis [15] by incorporating the physics and geometry priors of imaging model to build a unified deep learning framework. Lin *et al.* [13] presented a dual-domain network for metal artifact reduction in CT imaging with a differentiable Radon inversion layer connecting sinogram and image domains. Würfl *et al.* [14] mapped filtered back-projection to neural networks and introduced a cone-beam back-projection layer. Despite the integration of physics and geometry priors, these methods still follow the same methodology of training the network to learn the mapping from measurements to image. In contrast, our method uses the network to learn a continuous implicit neural representation of entire image by mapping spatial coordinates to corresponding intensity values. This is fundamentally different from previous deep learning-based reconstruction approaches.

### B. Implicit Neural Representation Learning

In recent years, the introduction of neural representations has completely changed representation and generation of images in deep learning [17]-[22]. First, Eslami *et al.* [17] introduced a generative query network to learn the scene representation from input images taken from different viewpoints, and used this representation to predict the image from unobserved viewpoints. Then, Sitzmann *et al.* [18] proposed a scene representation network to represent scenes as continuous functions that map world coordinates to local feature representation and formulate the image formation as a differentiable ray-marching process. Mildenhall *et al.* [19] used the fully-connected network to represent scenes and optimized neural radiance fields to synthesize novel views of scenes with complicated geometry and appearance. This year, Researchers have further improved the algorithms of implicit neural representation and extended them to broader applications. In [20] it was showed that Fourier feature mapping enabled MLP to learn high-frequency functions when using MLP to represent complex objects or scenes, while [21] demonstrated that periodic activation functions helped to represent complex signals and their derivatives with fine details. Moreover, Martin-Brualla *et al.* [22] extended neural radiance fields [19] to address real-world problems and enabled synthesis of novel views of complex scenes from unstructured internet photo collections with variable illumination.

Although much attention has been attracted for neural representation in natural images or photographic images, few studies have been done in the medical domain. In [19] some early discussions of CT and MRI reconstruction were included in one of the validation experiments, but the reconstructed image qualities are far from satisfactory. A work by Sun *et al.* on arXiv [23] is relevant. However, it only focused on using MLP to represent the measurement field instead of image, and then relying on other existing reconstruction methods to obtain reconstructed images. Our method aims to learn and optimize the neural representation for the entire image, and can directly reconstruct the target image by incorporating the forward model of imaging system.

### C. Image Prior Embedding

Deep neural networks have shown excellent performance for image generation and reconstruction, due to their ability to learn image priors from a large set of example images during training [9]. In addition, Ulyanov *et al.* [24] showed that the network structure itself can also capture low-level image statistics priors, where a randomly-initialized neural network given noise inputs can be used as a handcrafted image prior for inverse problems. Moreover, Gandelsman *et al.* [25] used multiple deep image prior networks for image decomposition in various applications. In this work, we go beyond leveraging such image priors through optimization in the function space of network's parameters. In particular, we also take advantage of another image prior unique in medical domain. In medical imaging it is common that one patient may have multiple imaging scans over time for the purpose of treatment assessment, or for image-guided interventions. Although the images are taken at different subject states, earlier scanned images can still provide useful prior knowledge of the patient anatomy. Our neural representation method proposes a simple yet effective way to embed this prior information and facilitate the reconstruction of the new images.

## III. METHOD

Fig. 1(B) illustrates the basic concept of the proposed implicit neural representation learning with prior embedding (NeRP) for image reconstruction. NeRP contains three modules to obtain the final reconstruction images. First, a prior image from earlier scan of the same subject is embedded as the implicit neural representation by encoding the entire spatial image field into the network's parameters. Specifically, the network is optimized to seek the continuous function that could precisely map the spatial coordinates to corresponding intensity values in the prior image. Next, using the prior-embedded network as the initialization, we aim to learn the neural representation of the target reconstruction image from the subsampled measurements of an unknown subject without any ground truth, as shown in Fig. 1(B). The differentiable forward model corresponding to the imaging system (e.g. Radon transform for CT imaging or Fourier transform for MRI imaging) is integrated to bridge between image space and sensor space. In this way, the network is optimized in the continuous function space of the network's parameters, with the constraints of the subsampled measurements from the unknown subject. Finally, the reconstructed image can be obtained by inferring the trained network across all the spatial coordinates in the image field.

### A. Problem Formulation

To formulate the inverse problem for computational image reconstruction, the forward process of imaging system can be modeled as:

$$y = Ax + e \quad (1)$$

where $x$ is the image of the unknown subject while $y$ is the sampled sensor measurements. Matrix $A$ represents the forward model of the imaging system, and $e$ is the acquisition noise.

Image reconstruction aims to recover the computational image $x$ of the unknown subject, given the measurements $y$ from sensors. In the sparsely sampled image reconstruction problem, the measurements $y$ are undersampled in sensor space for either accelerated acquisition, as in MRI, or reduction of radiation, as in CT. The inverse problem for sparse sampling is ill-posed, and is typically formulated as an optimization problem with regularization:

$$x^* = \underset{x}{\operatorname{argmin}}\ \mathcal{E}(Ax, y) + \rho(x) \quad (2)$$

where $\mathcal{E}(Ax, y)$ is the data term, which measures the errors between $Ax$ and $y$, so as to guarantee the data consistency with the sensor measurements. Function $\mathcal{E}$ can be different distance metrics such as L1 or L2 norm. $\rho(x)$ is the regularizer term characterizing the generic image prior. The regularizer $\rho(x)$ can be determined in many different ways to capture various image characteristics. For example, total variation of the image enforces smoothness, while sparsity in a transform domain is used in compressed sensing.

### B. Neural Representation for Image

In implicit neural representation learning, the image is represented by a neural network as a continuous function. The network $\mathcal{M}_\theta$ with parameters $\theta$ can be defined as:

$$\mathcal{M}_\theta : c \to v \quad with \quad c \in [0,1)^n,\ v \in \mathbb{R} \quad (3)$$

where the input $c$ is the normalized coordinate index in the image spatial field, and the output $v$ is the corresponding intensity value in the image. The network function $\mathcal{M}_\theta$ maps coordinates to the image intensities, which actually encodes the internal information of entire image into the network parameters. Thus, network structure $\mathcal{M}_\theta$ with the parameters $\theta$ is also regarded as the neural representation for the image. Note that, theoretically, a random image in any modality or in any dimension $x \in \mathbb{R}^n$ can be parameterized by the network using this method. Below we introduce the specific network structure used in our method.

#### 1) Fourier Feature Embedding

Since Fourier features are shown to be effective for networks to learn high-frequency functions [20], we use a Fourier feature mapping $\gamma$ to encode the input coordinates $c$ before applying them to the coordinate-based network. Thus, the encoded coordinates are:

$$\gamma(c) = [\cos(2\pi B c), \sin(2\pi B c)]^T \quad (4)$$

where matrix $B$ represents the coefficients for Fourier feature transformation. Following [20], entries of matrix $B$ are sampled from Gaussian distribution $\mathcal{N}(0, \sigma^2)$, where $\sigma$ is a hyperparameter characterizing the standard deviation of the prior distribution. After the Fourier feature embedding, the input to the network $\mathcal{M}_\theta$ is the encoded coordinates $\gamma(c)$.

#### 2) Multi-Layer Perceptron Network

The network $\mathcal{M}_\theta$ is implemented by a deep fully-connected network or multi-layer perceptron (MLP). The coordinate-based MLP parameterizes the continuous function to represent the entire image. This function is defined by the network structure as well as the network parameters. In the next section, we will describe in detail how to obtain the network parameters through optimization. For the network structure, the model depth and width of MLP are hyper-parameters, characterizing the representative capability of the MLP model. Moreover, we use the periodic activation functions in our MLP model after each fully-connected layer, which are demonstrated to effectively represent fine details in signals [21].

*C. NeRP for Sparsely Sampled Image Reconstruction*

Next, we introduce how the proposed implicit neural representation learning with prior embedding (NeRP) is used to solve image reconstruction problem. The goal is to recover the image $x$ of the target subject, given corresponding sparsely sampled measurements $y$ and a prior image $x^{pr}$. Note that $x^{pr}$ and $x$ are different scans for the same subject, but at different time points. These capture the changing state of the subject such as the tumor progression for monitoring patient response to therapy.

*1) Prior Embedding*

In the first step, we embed the prior image $x^{pr}$ into the network. We use the coordinate-based MLP $\mathcal{M}_\phi$ introduced in Sec.III.B to map the spatial coordinates to corresponding intensity values in prior image $x^{pr}$. That is, $\mathcal{M}_\phi: c_i \to x_i^{pr}$, where $i$ denotes the coordinate index in image spatial field. Given all the coordinate-intensity pairs in prior image $\{c_i, x_i^{pr}\}_{i=1}^N$ with a total of $N$ pixels in the image, the randomly-initialized MLP is optimized based on the objective:

$$\phi^* = \underset{\phi}{\mathrm{argmin}}\, \frac{1}{N}\sum_{i=1}^{N}\|\mathcal{M}_\phi(c_i) - x_i^{pr}\|_2^2 \quad (5)$$

After optimization, the internal information of prior image $x^{pr}$ is encoded into the MLP network $\mathcal{M}_{\phi^*}$ with the corresponding network parameters $\phi^*$. For clarity, we use $\mathcal{M}^{pr}$ to denote the prior-embedded MLP network, i.e. $x^{pr} = \mathcal{M}_{\phi^*} = \mathcal{M}^{pr}$.

*2) Network Training*

Given the prior-embedded MLP $\mathcal{M}^{pr}$ and measurements $y$, we train the network to learn the neural representation of the target image. Based on the formulation in Eq. (2), the unknown target image $x$ is parametrized by a coordinate-based MLP $\mathcal{M}_\theta$ with parameters $\theta$. Thus, the data term is defined as $\min_x \mathcal{E}(Ax, y) = \min_\theta \mathcal{E}(A\mathcal{M}_\theta, y)$, where the optimization in image space is transformed to the optimization in the space of MLP's parameters. Furthermore, we replace the explicit regularizer $\rho(x)$ by the implicit priors from both the prior image and the neural network parametrization (i.e. the internal information from prior image embedded in $\mathcal{M}^{pr}$ as well as the low-level image statistics prior captured by neural network parametrization $\mathcal{M}_\theta$), similar to the setting in [24]. More specifically, in our proposed approach, the specific implicit regularization means that all the pairs of coordinate-intensity share the same underlying continuous function that is parameterized by the MLP network. Besides having an initializer by embedding prior image as network weights, the implicit regularization captured by the network parametrization (i.e. the same continuous function parametrization) also influences the entire optimization process, which is realized through the neural representation learning. Thus, the optimization objective in Eq. (2) can be formulated as follows:

$$\theta^* = \underset{\theta}{\mathrm{argmin}}\, \mathcal{E}(A\mathcal{M}_\theta, y; \mathcal{M}^{pr}), \quad x^* = \mathcal{M}_{\theta^*} \quad (6)$$

The network $\mathcal{M}_\theta$ is trained by minimizing the L2-norm loss, which is initialized by the prior-embedded network $\mathcal{M}^{pr}$.

$$\theta^* = \underset{\theta}{\mathrm{argmin}}\, \|A\mathcal{M}_\theta - y\|_2^2, \quad x^* = \mathcal{M}_{\theta^*} \quad (7)$$

Note that forward model $A$ is adapted to the corresponding imaging system, such as Radon transform for CT imaging and Fourier transform for MRI imaging. The operation $A$ is differentiable, which enables training the network $\mathcal{M}_\theta$ in an end-to-end fashion.

*3) Image Inference*

Finally, after the network is well trained, the reconstruction image can be generated by inferring the trained network across all the spatial coordinates in the image field. That is: $x^*: \{c_i,\ \mathcal{M}_{\theta^*}(c_i)\}_{i=1}^N$, where $i$ denotes the coordinate index in image spatial field. This is denoted in short as $x^* = \mathcal{M}_{\theta^*}$ in Eqs. (6) and (7). Filling the intensity values at all the coordinates in image grid constitutes the final reconstruction image $x^*$.

*D. Technical Details of NeRP*

In our implementation, we construct an 8-layer MLP network with a width of 256 neural nodes for CT reconstruction, where each fully-connected layer is followed by the periodic activation function [21] except for the last layer. For MRI reconstruction, we increase the MLP width to 512 layers. We will discuss and analyze the influence of different network structures in next Section. The Fourier feature embedding [20] size is 256, where the hyper-parameter for the standard deviation of the coefficient's Gaussian distribution is set as 3 for MRI reconstruction and 4 for CT reconstruction. For prior embedding, the training objective in Eq. (5) is optimized by the Adam optimizer with a learning rate of 0.0001. The total training iterations are 1000 for 2D images and 2000 for 3D images. Next, given the prior-embedded MLP as the initialization, the reconstruction network is trained by optimizing the objective in Eq. (7) using the Adam optimizer with a learning rate of 0.00001. Usually we train 1000 iterations for 2D images and 2000 iterations for 3D images. We implemented our networks using PyTorch [26]. For the differentiable forward model $A$, the Radon transform or forward projection operation for CT imaging is realized by using Operator Discretization Library (ODL) [27]. The non-uniform Fast Fourier Transform (NUFFT) for MRI imaging is implemented based on the *torchkbnufft* package [28].

IV. EXPERIMENTS AND RESULTS

To evaluate the proposed NeRP method, we conducted experiments for 2D/3D CT and MRI image reconstruction with sparsely sampling. For CT image reconstruction we assume 20 projections equally distributed across a semi-circle. We compute parallel-beam projections for 2D CT and cone-beam projections for 3D CT. For MRI image reconstruction, 40 radial a public dataset for brain tumor progression [29] [30]. This spokes are sampled in k-space with golden angle as the angular interval, as shown in Fig. 1(B). Beyond sparsely-sampled measurements data, a prior image from an earlier scan is also given. Since the prior image and reconstruction image are of the same patient at different time points, the prior image can provide useful information about the patient's anatomic structure while still allowing crucial structural and functional differences such as tumor or lesion changes, as shown in Fig. 1(B). We will show the experimental results applying NeRP for 2D/3D CT and MRI image reconstruction with various image contrasts and at various anatomical sites.

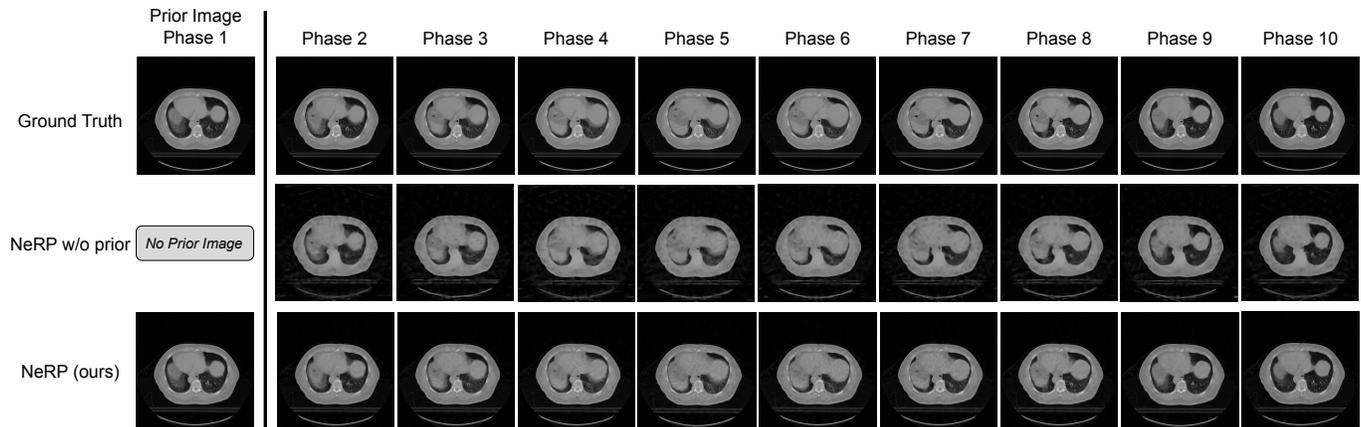

Fig. 2. Results of 2D CT image reconstruction for pancreas 4D CT data using 20 projections. The first row shows the ground truth cross-sectional 2D slices at the same location over 10 phases in the pancreas 4D CT, where each column demonstrates one phase respectively. The final row shows the reconstruction images at different phases respectively by using the proposed NeRP method, where the phase-1 image is used as the prior image for reconstructing the images in phase 2 ~ 10. For comparison, the second row shows the reconstruction results without using the prior embedding.

*A. Datasets*

*1) Pancreas 4D CT data*

For CT image reconstruction, we collected a pancreas 4D CT data from a clinical patient. The 4D CT data is a temporal scan with a sequential series of 3D CT images over a respiratory cycle. Due to respiratory motion there is continuous motion in the CT images at different time points. In the first row of Fig. 2, we show the cross-sectional images at the same location from the 10 phases (time-points) of the CT images in the 4D CT. The continuous structural changes can be observed over the 10 temporal phases. In the experiments we use phase 1 as the prior image to reconstruct the CT images at the subsequent phases.

*2) Head and Neck CT and Lung CT data*

To further validate the generalization of the proposed NeRP algorithm, we collected two clinical patient cases including a head and neck CT case and a lung CT case. For each case there are two longitudinal 3D CT images scanned for the same patient at time points during treatment with radiation therapy. The goal is to follow tumor volume to assess response to therapy. In the data preprocessing, we firstly conduct rigid image registration to align the two CT images at the same position. Then, we use NeRP to reconstruct the latter 3D CT image while using the earlier 3D CT image as the prior image.

*3) Brain Tumor Progression MRI data*

For MRI image reconstruction we conducted experiments on dataset includes MRI images from 20 subjects with primary newly diagnosed glioblastoma. The patients were treated with surgery and standard concomitant chemo-radiation therapy (CRT) followed by adjuvant chemotherapy [30]. For each patient, there are two MRI exams included, which were within 90 days following CRT completion and at tumor progression. Thus, the tumor changes can be clearly observed by comparing the two MRI exams of the same patient at different time points. In addition, each MRI exam contains multi-contrast MRI images including T1-weighted, and contrast-enhanced T1-weighted (T1c), T2-weighted, FLAIR. In our experimental setting, we set the first MRI exam as the prior image and aim to reconstruct the MRI image in the second exam. This is tested for different MR image constrasts respectively.

*B. Experiments on 2D CT Image Reconstruction*

In Fig. 2, we show the 2D CT reconstruction results for phase 2 to phase 10 using the proposed NeRP algorithm by using phase 1 as prior image. Note that 2D CT images are the cross-sectional slices at the same location extracted from the corresponding 3D CT images at each phase. After pre-processing, 2D CT images are all resized to $256 \times 256$. For comparison, we demonstrate reconstruction results for "NeRP w/o prior", where no prior image is used and the network is randomly initialized for training in Fig. 1(B). From the results, we can see that NeRP can reconstruct high-quality images with clear anatomic structures, sharp organ boundary and high-contrast soft tissue and bones. More importantly, the reconstructed images can precisely capture the continuous changes with fine detail over different phases, although the same prior image is used and only sparse projections are sampled for reconstructing the target image in each phase. Note that neither the method of "NeRP w/o prior" nor "NeRP" requires a large amount of images from different patients for training the network, since both methods are based on neural representation learning. The difference is in the prior embedding. By comparing to the results of "NeRP w/o prior", we see prior image embedding benefits the reconstruction results in all phases with increased image sharpness and reduced noise. Therefore, we conclude that the prior embedding can provide useful prior knowledge that is critical to precisely reconstruct high-quality 2D CT images with sparse sampling. Moreover, we see that the proposed NeRP algorithm can reconstruct reliable images that precisely capture the small structural changes in the patient's anatomy.

### C. Experiments on 3D CT Image Reconstruction

To evaluate the effectiveness of NeRP in a higher-dimensional reconstruction task, we conducted experiments for 3D CT image reconstruction. In the first experiment on the pancreas 4D CT data, we use the entire phase-1 3D CT as the prior image and aim at reconstructing the 3D CT image at phase 6 with image size of $128 \times 128 \times 40$ after image cropping and resizing. Note that phase 1 and phase 6 are exactly the inhalation and exhalation phases during this 4D CT scan, which have the largest structural difference. Figure 3 shows the prior images and ground truth images of reconstruction target in the first two rows, where each column shows the cross-sectional image of the entire 3D volume. There are clear anatomic structural differences between these two phases as highlighted by the red boxes. The final row demonstrates the reconstructed 3D volumetric image using the proposed NeRP algorithm. We can see that the reconstructed image captures the correct anatomic structures in the target phase as indicated by the red boxes, with high image quality and image contrast.

TABLE I
RESULTS OF 3D CT IMAGE RECONSTRUCTION
USING 5 / 10 / 20 PROJECTIONS ON DIFFERENT ANATOMICAL SITES

| Methods | Pancreas CT | HeadNeck CT | Lung CT |
|---|---|---|---|
| Projections = 10 | | | |
| FBP | 17.95 / 0.461 | 23.05 / 0.653 | 21.49 / 0.597 |
| GRFF [20] | 28.07 / 0.855 | 29.38 / 0.864 | 27.80 / 0.835 |
| NeRP w/o prior | 28.88 / 0.850 | 30.40 / 0.858 | 30.98 / 0.880 |
| NeRP (ours) | **37.66 / 0.981** | **36.92 / 0.976** | **32.73 / 0.941** |
| Projections = 20 | | | |
| FBP | 18.23 / 0.610 | 23.42 / 0.750 | 21.74 / 0.717 |
| GRFF [20] | 29.27 / 0.893 | 32.56 / 0.931 | 32.75 / 0.935 |
| NeRP w/o prior | 32.41 / 0.927 | 32.59 / 0.920 | 32.86 / 0.929 |
| NeRP (ours) | **39.06 / 0.986** | **38.81 / 0.985** | **36.52 / 0.972** |
| Projections = 30 | | | |
| FBP | 18.31 / 0.650 | 23.54 / 0.773 | 21.83 / 0.7443 |
| GRFF [20] | 31.53 / 0.932 | 32.34 / 0.927 | 33.13 / 0.942 |
| NeRP w/o prior | 33.88 / 0.953 | 33.53 / 0.942 | 33.97 / 0.951 |
| NeRP (ours) | **39.65 / 0.987** | **39.50 / 0.987** | **37.66 / 0.980** |

Evaluation metric: PSNR / SSIM values are reported.
PSNR (dB), peak signal noise ratio; SSIM, structural similarity.

For comparison, we also conducted experiments to show the results of other reconstruction methods. First, we show the reconstruction results for "NeRP w/o prior" as an ablative study by removing the prior embedding. Comparing the image quality, we can see that the prior embedding effectively contributes to reconstructing high-quality images with sparse sampling. Moreover, we compare with the analytic reconstruction algorithm using filtered back projection (FBP).

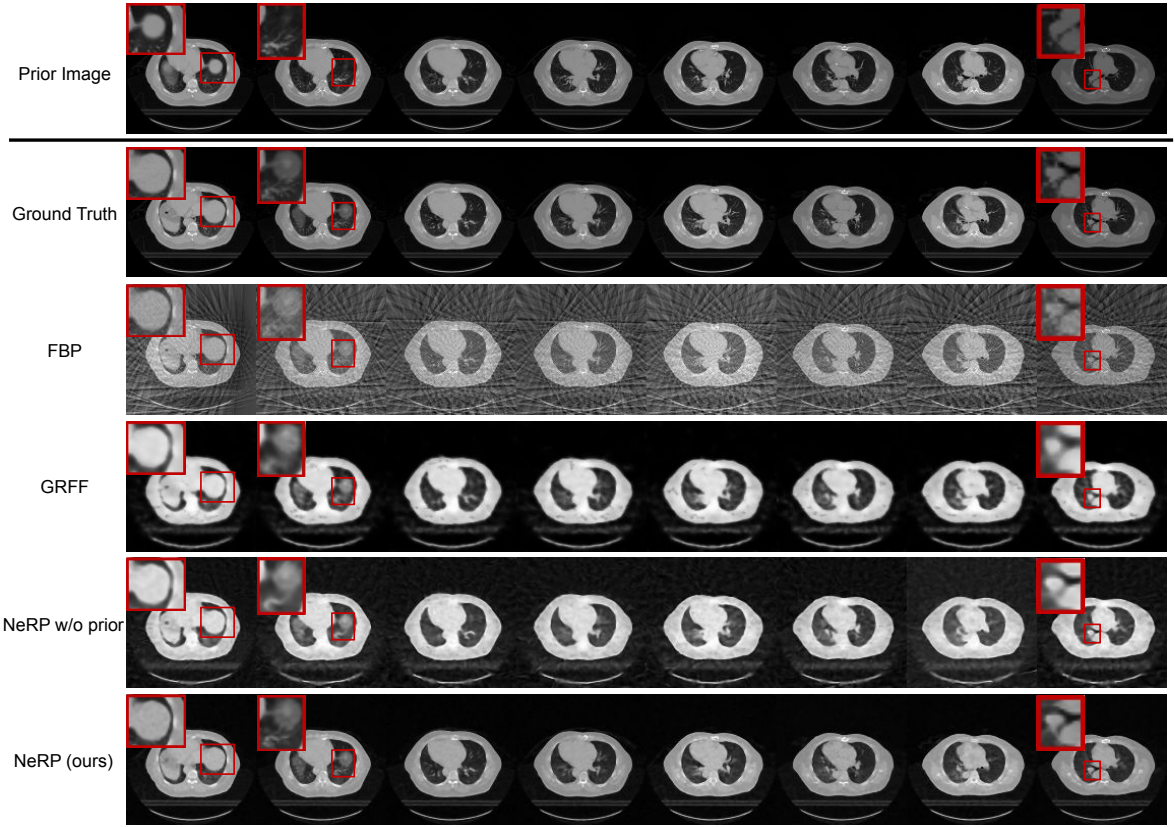

Fig. 3. Results of 3D CT image reconstruction for pancreas 4D CT data using 20 projections. The first and second rows show the prior 3D CT (phase 1) and the ground truth of target 3D CT (phase 6) image, where each column demonstrates cross-sectional slices of the 3D volume. The final row shows the reconstructed 3D CT images by using the proposed NeRP method. For comparison, the second to fourth rows show the reconstruction results of FBP method, GRFF method [20] and NeRP without using prior embedding. The zoom-in red boxes highlight the difference in anatomic structure for comparison.

The back-projected operation adjoint to the cone-beam projection in the forward model can reconstruct the 3D image from the given 2D projections with filter correction. As shown in Fig. 3, FBP introduces severe streaking artifacts due to the sparsely sampling measurements. Usually, the FBP algorithm requires hundreds of sampled projections to satisfy the Shannon-Nyquist theorem in order to obtain artifact-free images. However, with the implicit image priors captured from deep network and prior embedding, the proposed NeRP can overcome this limit and achieve artifact-free high-quality images with only sparsely sampled projections. In addition, we also compare with a relevant previous work [20], which also attempts to represent medical images by network-based continuous functions. We compare with the best method reported in [20] using Gaussian random Fourier feature (GRFF) and implement the method according to the technical details in [20]. Compared with GRFF, our proposed NeRP uses a different network architecture, and more importantly, introduces the prior embedding for learning implicit neural representations. As shown in Fig. 3, in the same 3D CT image reconstruction task, our method obtains reconstructions with better image quality than GRFF.

Going beyond 4D CT data, we also evaluated the clinical radiation therapy patient data with both head and neck CT and lung CT. The quantitative results for 3D CT reconstruction evaluated by PSNR and SSIM metrics are reported in Table I on different anatomic sites including pancreas CT, head and neck CT and lung CT with all comparison methods. To evaluate the effectiveness under different settings, we also compare the reconstruction results with different numbers of projections. Our proposed NeRP method achieves the best performance in either metric for all the 3D CT image cases with 10 / 20 / 30 projections respectively, outperforming all the other methods that don't use prior image embedding. Reconstructed images for the head-and-neck CT case are shown in Fig. 4. The prior image (former scan) and the ground truth target image (latter scan) are shown in the first two rows. The reconstructed 3D CT image using our NeRP method are demonstrated in the last row, which successfully captures the accurate anatomic structure and fine details different from prior image as pointed out by the red boxes. Besides, comparing with other methods including FBP, GRFF and NeRP w/o prior, our method is able to reconstruct higher-quality images with sharper organ boundaries, higher-contrast bone regions and reduced noise and artifacts. More importantly, we show that NeRP is a general reconstruction methodology that can be applied to different body sites across different patients, as shown in these two cases for head-and-neck and lung. This demonstrates the potential capability of NeRP for practical clinical application.

### D. Experiments on 2D MRI Image Reconstruction

We conducted experiments to evaluate the proposed method for MRI image reconstruction. We aimed to reconstruct 2D MRI images with sparsely sampled frequency space (k-space) data by using a radial sampling pattern for data acquisition, which is widely used in clinical MRI. The 2D NUFFT for radial sampling is used as the forward model to compute and sample k-space data as shown in Fig. 1(B). In addition to subsampled k-space measurements, we also assume that a prior image from a previous scan is available. In the brain tumor regression dataset, we use the first MRI exam as the prior image to reconstruct the MRI image in the second exam. After preprocessing, 2D MRI images are all resized to $256 \times 256$.

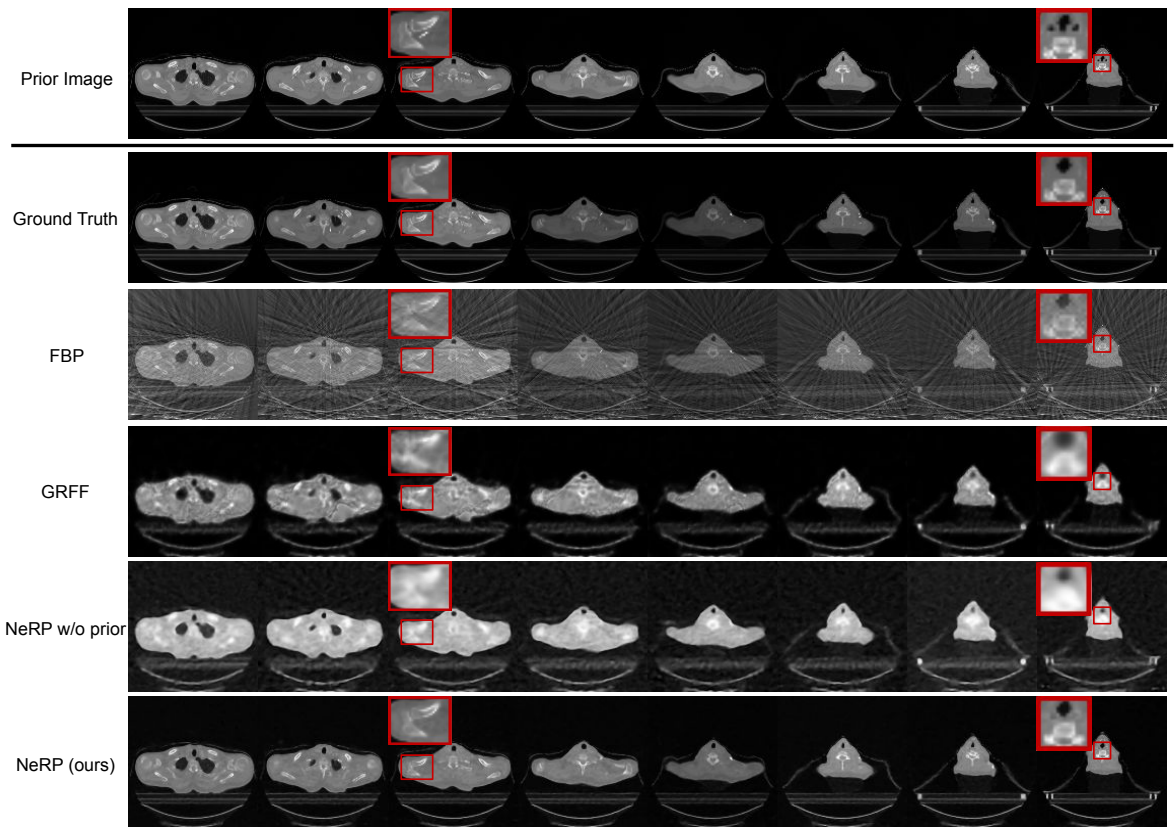

Fig. 4. Results of 3D CT image reconstruction for longitudinal head-and-neck CT case using 20 projections. The first and second rows show the prior 3D CT (former scan) and the ground truth target 3D CT (latter scan) image, where each column demonstrates cross-sectional slices of the 3D volume. The final row shows the reconstructed 3D CT images by using the proposed NeRP method. For comparison, the second to fourth rows show the reconstruction results of FBP method, GRFF method [20] and NeRP without using prior embedding. The zoom-in red boxes highlight the difference in anatomic structure for comparison.

Figure 5 demonstrates 2D MRI reconstruction results for multi-contrast MR images. The first and second rows show the prior image (the first exam) and the ground truth target image (the second exam) for one randomly selected patient with four image contrasts (T1, T1c, T2, FLAIR). For each constrast, there are two cross-sectional 2D MRI images as two separate reconstruction cases. Comparing the prior images and target images, we can clearly see the tumor progression with the changed shape and size in these two exams. The reconstruction results for NeRP and NeRP w/o prior are shown in the third and fourth row, respectively. For better visualization and comparison, we zoom in and crop the sub-image of the corresponding tumor regions of the ground truth and reconstruction images. From the comparison, it is seen that the reconstructed images from sparsely subsampled k-space data can accurately capture the fine detailed structures especially in the tumor region, which differs from that in the prior image. It is clear that the prior embedding can help to reconstruct high-quality images from sparsely sampled k-space data in the second exam by efficiently exploiting prior knowledge for the same patient. More importantly, the proposed reconstruction algorithm can be easily generalized to different MR image contrasts, which indicates the potential for other practical applications.

### E. Experiments on 3D MRI Image Reconstruction

Using the same dataset with brain tumor regression, we further evaluated the 3D MRI image reconstruction. In this case, the entire 3D MRI volume in the first exam is used as the prior image in order to reconstruct the 3D MRI image in the second exam for the same patient. The forward model is the 3D NUFFT to compute and sample 3D k-space data. The whole learning framework of NeRP is similar to that of 2D MRI reconstruction except for using a 3D coordinate index. In pre-processing, all the 3D MRI images are cropped and resized to $128 \times 128 \times 24$.

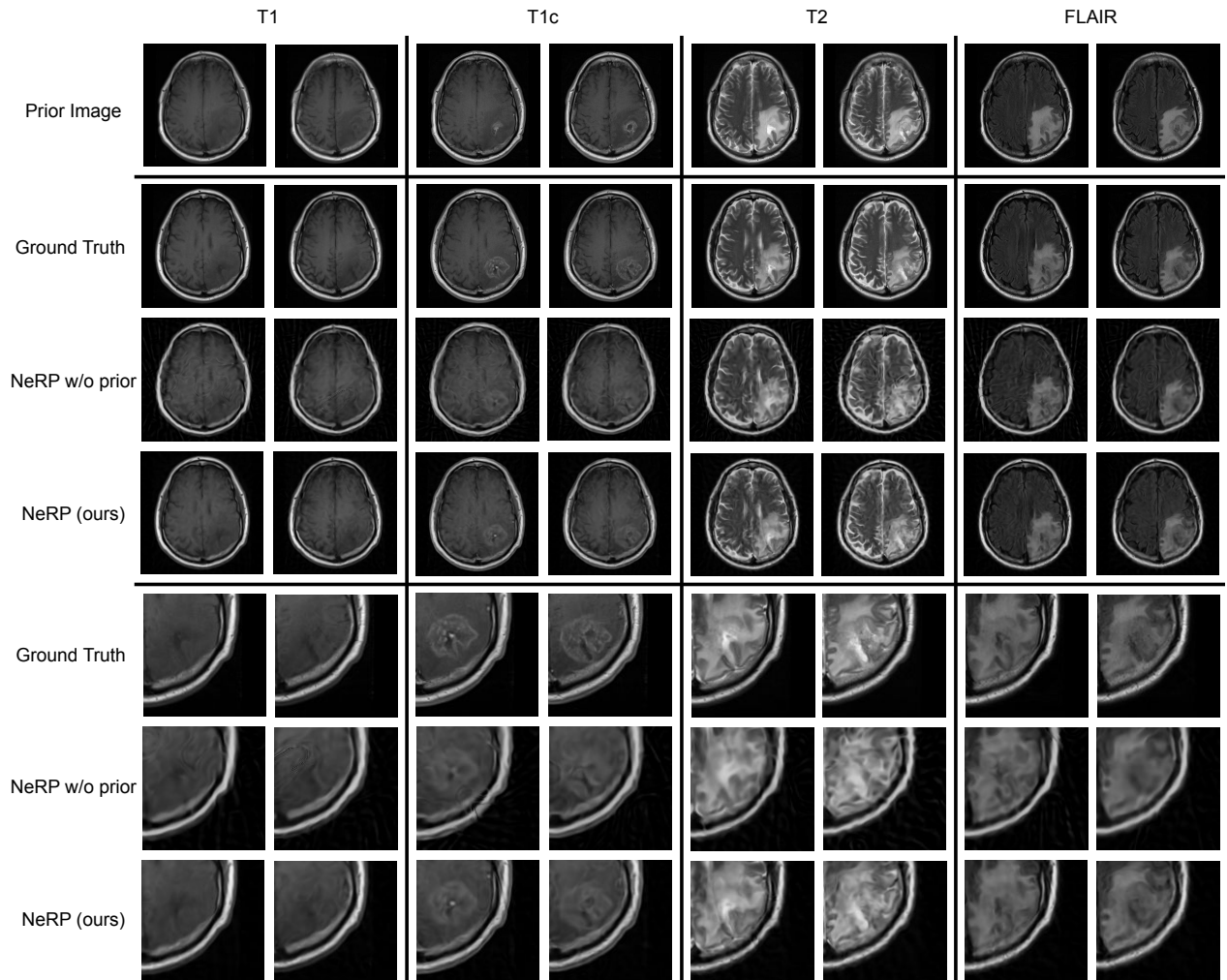

Fig. 5. Results of 2D MRI image reconstruction for multi-contrast MRI data using 40 radial spokes. The first and second rows show the 2D MRI images of the prior image (first exam) and the target image (second exam) for four image contrasts (T1, T1c, T2, FLAIR). The third and fourth rows show the reconstruction results by using the proposed NeRP method as well as the ablation study without using prior embedding. By zooming in the tumor regions, the last three rows show the cropped sub-images of the tumor regions corresponding to the images in the second, third, and fourth rows, respectively.

Quantitative results of 3D MRI image reconstruction evaluated by PSNR and SSIM metrics are reported in Table II for different image contrasts including T1, T1c, T2 and FLAIR. We compare the results of different reconstruction methods with 30 / 40 / 50 sampled radial spokes, respectively. From the Table II, our NeRP method achieves better performance than other methods without using prior image for all the image contrasts. Reconstructed 3D MRI images for FLAIR contrast are demonstrated in Fig. 6. We first show the prior 3D MRI image (first exam) and the ground truth 3D MRI image (second exam), where each column demonstrates cross-sectional slices of the 3D volume. The final row shows the reconstructed 3D MRI images from the proposed NeRP algorithm. The reconstruction results indicate that our method is able to reconstruct the precise changes in brain tumor region (pointed out by the red boxes) even with sparsely sampled k-space data, which is crucial for clinical diagnosis and cancer treatment. Moreover, for comparison, we also use an analytic reconstruction method with the Adjoint NUFFT operator to recover the data in image space with density compensation. We can see the reconstructed images using adjoint NUFFT have severe streaking artifacts due to the sparse down-sampling of k-space data. In addition, we also compare with GRFF method [20] and NeRP w/o prior as ablative study. The comparison shows that our method can not only reconstruct more accurate tumor structures as pointed out by the red boxes, but also achieves better image quality such as sharper anatomic boundaries and higher soft tissue contrast. All of these results demonstrate the effectiveness and superiority of the proposed NeRP algorithm for 3D MRI image reconstruction.

TABLE II
RESULTS OF 3D MRI IMAGE RECONSTRUCTION
USING 30 / 40 / 50 RADIAL SPOKES FOR DIFFERENT IMAGE CONTRASTS

| Methods | T1 | T1c | T2 | FLAIR |
|---|---|---|---|---|
| Spokes = 30 | | | | |
| Adjoint NUFFT | 20.91 / 0.63 | 21.68 / 0.63 | 19.55 / 0.57 | 19.77 / 0.58 |
| GRFF [20] | 27.98 / 0.90 | 27.67 / 0.88 | 25.66 / 0.85 | 25.98 / 0.86 |
| NeRP w/o prior | 27.49 / 0.85 | 27.82 / 0.87 | 25.91 / 0.85 | 26.87 / 0.88 |
| NeRP (ours) | **28.43 / 0.90** | **29.06 / 0.92** | **26.86 / 0.90** | **27.52 / 0.90** |
| Spokes = 40 | | | | |
| Adjoint NUFFT | 21.30 / 0.66 | 22.05 / 0.67 | 20.17 / 0.62 | 20.23 / 0.61 |
| GRFF [20] | 28.18 / 0.90 | 28.11 / 0.89 | 25.67 / 0.85 | 25.99 / 0.86 |
| NeRP w/o prior | 29.70 / 0.92 | 29.29 / 0.91 | 27.59 / 0.91 | 27.54 / 0.90 |
| NeRP (ours) | **31.75 / 0.96** | **30.53 / 0.94** | **28.73 / 0.93** | **29.07 / 0.93** |
| Spokes = 50 | | | | |
| Adjoint NUFFT | 21.40 / 0.68 | 22.26 / 0.69 | 20.42 / 0.64 | 20.49 / 0.64 |
| GRFF [20] | 28.50 / 0.91 | 27.59 / 0.88 | 25.23 / 0.85 | 25.90 / 0.87 |
| NeRP w/o prior | 30.65 / 0.94 | 29.26 / 0.91 | 28.40 / 0.92 | 27.68 / 0.90 |
| NeRP (ours) | **32.55 / 0.96** | **31.37 / 0.95** | **30.13 / 0.95** | **30.02 / 0.94** |

Evaluation metric: PSNR / SSIM values are reported.
PSNR (dB), peak signal noise ratio; SSIM, structural similarity.

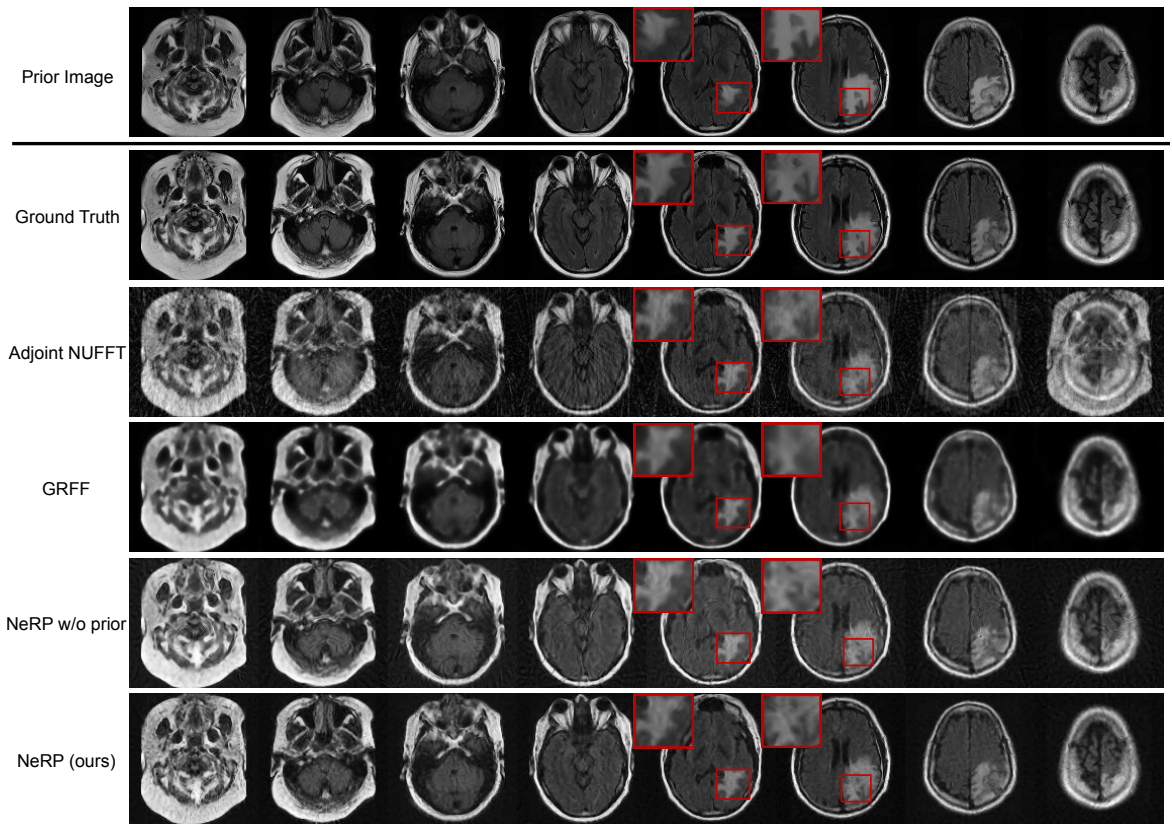

Fig. 6. Results of 3D MRI image reconstruction for FLAIR contrast using 40 radial spokes. The first and second rows show the prior 3D MRI (first exam) and ground truth of target 3D MRI (second exam) image, where each column demonstrates cross-sectional slices of the 3D volume. The final row shows the reconstructed 3D MRI image by using the proposed NeRP method. For comparison, the second to fourth rows show the reconstruction results of Adjoint NUFFT method, GRFF method [20] and NeRP without using prior embedding as ablative study. The zoom-in red boxes highlight the difference in brain tumor region for comparison.

TABLE III
ANALYSIS OF NETWORK STRUCTURE
FOR 3D CT / MRI IMAGE RECONSTRUCTION

| Network Structure | | Pancreas CT | T1 MRI |
|---|---|---|---|
| Width=512 | 4 Layers | 38.04 / 0.981 | 29.93 / 0.933 |
| | 6 Layers | 38.83 / 0.984 | 31.23 / 0.950 |
| | 8 Layers | 37.44 / 0.977 | **31.75 / 0.956** |
| Width=256 | 8 Layers | **39.06 / 0.986** | 31.10 / 0.948 |
| | 16 Layers | 34.04 / 0.953 | 30.98 / 0.945 |
| | 20 Layers | 34.04 / 0.955 | 30.23 / 0.937 |

Evaluation metric: PSNR / SSIM values are reported.
PSNR (dB), peak signal noise ratio; SSIM, structural similarity.

### F. Analysis of Network Structure

For the proposed NeRP algorithm, one important issue is to set a proper network structure for the MLP backbone. The MLP network parameters serve as the variables to expand the function space for network optimization and seeking the optimal reconstructed images. The number of network parameters is related to the depth and width of the MLP, i.e., the number of layers and the number of neurons in each layer. To analyze the influence of network structures, we conducted an ablation study to obtain the reconstruction results with changed MLP depth and width as shown in Table III. Here, the 3D pancreas CT image is reconstructed from 20 projections while the 3D T1 MRI image is reconstructed from 40 radial spokes. From Table III, we see that the reconstruction results are not very sensitive to the change of network depth or width, which indicates the proposed NeRP algorithm is robust to the specific choice of network structure. In experiments, we also observe that training the MLP model could be more difficult with more layers, where the insufficient optimization may cause worse reconstruction results.

### G. Analysis of Sparse Sampling Ratio

To better analyze the influence of the sparse sampling ratio, we use NeRP to reconstruct 3D CT and MRI images with different number of sampled projections or radial spokes. Fig. 7 and Fig. 8 show the PSNR and SSIM of reconstructed CT and MRI images with increasing sampling, respectively. The curves show that more samples in the measurements field can always reconstruct more precise structures with better image quality. For CT images, sampling around 20 projections reaches the plateau when using NeRP reconstruction algorithm, while around 40 sampled radial spokes are required for MRI image reconstruction to achieve the best image quality. The analysis provides guidance for designing sensor acquisition methods for imaging system when using NeRP algorithm for image reconstruction.

## V. DISCUSSION

Although many existing methods have shown the advantages of deep learning models for medical image reconstruction [31], there are increasing concerns regarding the limitations of the current deep learning approaches. First, training deep neural networks is data-intensive and requires large-scale datasets. This may prevent many practical applications due to the difficulty of data collection. The limited availability of specific image modalities or images of rare diseases may make it difficult to acquire sufficient training data for deep learning modeling. In addition, there is a common concern about the robustness and reliability of deep reconstructed images. For example, Antun et al. [16] find that the small but significant structural changes in tumors or lesions may not be accurately captured in the deep reconstructed images. Finally, the generalization ability of deep networks is unclear. The trained deep networks may produce poor performance when generalized to data out of the training distribution, such as across different imaging modalities or contrasts, or across different anatomical sites. The model generalizability is related to not only the training data distribution but also the network structure. For example, it is a non-trivial task to directly transfer a deep network developed for MRI reconstruction to the task of CT reconstruction due to different sensor measurements fields. Because of these issues and limitations, new insights are needed in developing deep learning-based image reconstruction algorithms.

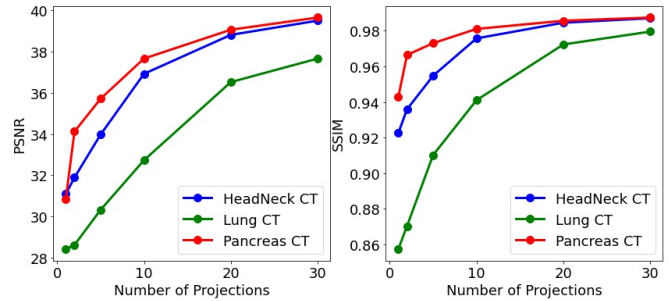

Fig. 7. Results of 3D CT image reconstruction (PSNR and SSIM) with different number of sampled projections.

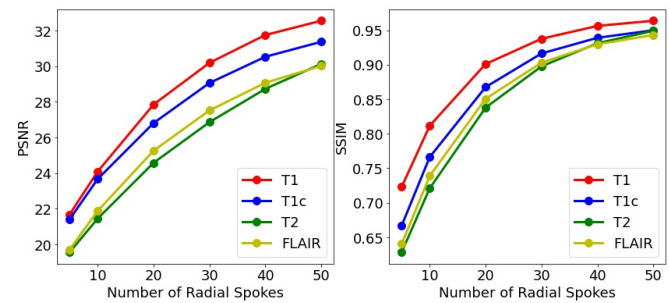

Fig. 8. Results of 3D MRI image reconstruction (PSNR and SSIM) with different number of sampled radial spokes.

The proposed NeRP method provides a new perspective to the problems of image reconstruction in both problem formulation and training process, which promises to overcome the shortcomings of previous deep reconstruction approaches. First, previous deep approaches to learn the mappings from sampled measurements to reconstructed images require a large-scale paired dataset (measurements, image) collected from different patients to train the model, which presents a bottleneck for practical applications of deep learning-based image reconstruction. The proposed method based on neural representation learning trains a patient-specific network by learning the coordinate-intensity mapping, where each pair of (coordinates, intensity) is from the same data sample. Thus, the proposed approach does not require humongous image datasets for model training and testing. In addition, from our experiments, the reconstructed images from NeRP are more robust and reliable, and can capture the small structural changes such as tumor or lesion progression. The implicit priors captured by network structure and the prior embedding can effectively incorporate the prior knowledge in the reconstruction process, which makes it possible for the network to capture and reconstruct the fine structural details in the resultant images. As a result of the previous two points, NeRP can be more easily generalized to different image modalities and contrasts, different anatomical sites, and different dimensionalities in the image reconstruction task. The relaxed requirements for training data make the method more transferrable and applicable across various applications. The proposed NeRP is a general methodology for medical image reconstruction with promising advantages over the existing deep reconstruction methods.

By embedding the image prior as network parameters, the proposed method is designed to exploit the anatomic information as the image prior. Given the overfitted nature of NeRP modeling, the similarity between the prior image and the image to be reconstructed is important to the ultimate success of the model. To a large extent, limiting the study to patient-specific NeRP image reconstruction is to ensure this similarity between the prior and the image to be reconstructed. It is interesting that, in patient-specific reconstructions, the proposed NeRP can accurately capture the structural difference between the prior image and the target image, as shown in the results in Figs. 2-6. In other words, the proposed method is not simply copying the prior image as reconstruction results, the prior-embedded network is sufficiently optimized under the constraints of the sparsely sampled measurements from the target image scanning. Thus, the reconstructed image can accurately match the target image.

The proposed method is a general approach of prior-embedded neural representation learning, which can be used for both medical and natural images. However, unlike in many medical applications in which the longitudinal prior images of the same patient are readily available for the prior embedding, in natural image applications, the longitudinal prior image of the same imaging object may not exist and how to find the best prior for natural image applications is not very clear. This is an interesting research question and worthy of further investigations in the future.

Reliability is a significant issue in image reconstruction. Previous works provide different solutions to address the convergence and reliability issue in deep learning-based sparse image reconstruction [32-34]. The proposed NeRP method has a few salient features. Firstly, there is no GAN-based training loss in the network optimization, which alleviates the enforcement for the reconstructed image to be realistic. Secondly, the network in the proposed approach is not trained from a population-based dataset collected from different patients, which mitigates the concern of model applicability when applied to a new patient. With prior image embedding from the same patient and implicit regularization captured by network structure, the network is optimized to learn a better representation of the patient to satisfy the constraints of given sparsely sampling with high fidelity. Therefore, the proposed reconstruction method is potentially more reliable for the specified applications.

Because of these unique characteristics as aforementioned, our method could be very valuable in medical clinical applications. In practical applications, for a new patient or subject, the model needs to be retrained. Since the proposed approach does not require abundant data for model training, re-training the model for a specific patient is computationally efficient and takes only minutes to complete in our experiments using a single Nvidia V100 GPU. This makes our method a good fit for real-world applications.

## VI. CONCLUSION

In this work, we propose a new deep learning-based medical image reconstruction methodology by learning implicit neural representations with prior embedding (NeRP), which efficiently incorporates the prior knowledge and learns to reconstruct the target image through implicit neural representations. Through the experiments for 2D/3D MRI and CT image reconstruction, we show that the proposed NeRP algorithm is able to provide high-quality reconstruction images even with sparsely sampled measurements data. The NeRP approach possesses a number of unique advantages: (1) requires no training data from external subjects for developing networks; (2) accurate reconstruction of small and detailed changes such as anatomic structure or tumor progression; (3) broad applicability to different body sites, different imaging modalities and different patients. For medical images, it is common that a patient is scanned multiple times for clinical diagnosis or treatment follow-up, for the purpose of treatment planning or monitoring of the changes in tumor volume before and after therapy. In a longitudinal image series, previous scans can provide useful prior knowledge for NeRP image reconstruction. The effectiveness of NeRP and advantages of prior embedding have been demonstrated in the extensive experiments.

CONFLICT OF INTEREST

L.S and L.X. are co-inventors on a patent application based on this work.

ACKNOWLEDGMENT

The authors acknowledge the funding supports from the Stanford Bio-X Bowes Graduate Student Fellowship of Stanford University, NIH/NCI 1R01CA227713 and 1R01CA256890.